\documentclass[12pt]{article}
\usepackage{amssymb}
\usepackage{epsfig}

\usepackage{graphicx}
\usepackage{amsmath}

\begin{document}
\date{}
\title{The Hierarchy Problem and the Safety in Colliders 
Physics\footnote{Talk given at 
``Les Rencontres de Physique de la Vallee d' Aoste'', La Thuile, 
March (2008).}}
\author{{\small Francesco Caravaglios} \\
{\small Dipartimento di Fisica, Universit\`{a} di Milano, Via Celoria 16,
 I-20133 Milano, Italy}
\\ {\small and}\\{\small INFN\ sezione di Milano}}
\maketitle

\begin{abstract}
The Standard Model correctly describes all interactions at (and below) the 
electroweak scale. 
However it does not explain the peculiar pattern of quark, lepton and neutrino 
masses. Also charge quantization is not understood. These are well known 
motivations to go beyond the Standard Model and to build a 
Grand Unified Theory. This extension has several  good predictions but the 
proton lifetime is huge compared to similar weak decays. This hierarchy problem
suggests two possible extensions of the standard quantum field theory: 
a non linear version of the Schroedinger functional equation and 
Third Quantization. We will make a comment on the safety of collider physics 
in the context of the non linear extension of QFT.
     
\end{abstract}

\section{Introduction}

The theory that describes the strong, electromagnetic and weak interactions is 
based on the gauge group $SU(3)\times SU(2)\times
U(1). $ 
The symmetry group is spontaneously broken and the gauge boson together with
 the 
matter fermions become massive.
If and only if the scalar field responsible for electroweak symmetry breaking 
is 
a  $SU(2)$ doublet with hypercharge -1/2 we get the well known relation  
\begin{equation}
\frac{M_{W}^{2}}{M_{Z}^{2}}=\cos ^{2}(\theta _{W})
\end{equation}
that relates the weak boson masses with the coupling constants in the
 interaction 
between  weak boson 
 and fermions.
Also charge quantization comes from this peculiar choice for the Higgs 
hypercharge, and this choice 
 is natural  in Grand Unified Theories as we 
will see later. 
The Standard Model gives a correct description of all forces that act at and 
below the weak scale. In fact it provides us with several theoretical
 predictions 
for all the observables  listed in Table 1.

Adding an extra (universal) $Z^{\prime }$ \cite{Chanowitz:2008ix} 
 or additional Higgs doublets does not significantly improve 
the fit of data; on the contrary these extensions of the Standard Model 
are strongly constrained by these data (Table 1). 
The the top mass obtained in this fit is in very good agreement with the
 direct 
experimental observation. The Higgs mass seems to be not very large, probably 
the Higgs  is lighter than the top. When the top mass is very heavy, as 
proven by experiments,
  the radiative corrections to the effective potential are large.
This  theoretical extrapolation of the standard theory to values of the Higgs 
 average  field much higher than the weak scale, shows that the value 
of 246 GeV 
 deduced from the weak boson masses is not a global minimum if  $M_H$ does not
 satisfy the inequality \cite{Sher:1993mf,Isidori:2007vm}
\begin{equation}
M_{H}\gtrsim 75+1.64(m_{t}-140)-3(\frac{\alpha _{s}-0.117}{0.007}).  
\label{eq1}
\end{equation}
This limit holds in the standard theory.
As we will see after, the effective potential is a theoretical extrapolation 
of the energy of the universe to quantum  physical states very far 
from the present universe that we observe, however we know that theories often 
have a wide validity region  that can often cover several order of magnitudes.
The validity of maxwell equations, as well as quantum mechanics have been
 proved 
in several extremely different experimental situations. 
If the effective potential of the standard theory has a validity extended 
over several order of magnitudes of the Higgs average field could be
 challenged 
not only by the limit   (\ref{eq1}) but also by the so called hierarchy 
problem 
that appears when the  group  $SU(3)\times SU(2)\times U(1)$ is embedded 
into a unified gauge group.
We mention the following arguments  that motivate us to embed the standard 
theory into a  grand unified theory.
The first motivation  is the charge quantization and the quantum numbers of 
the 
matter fermions. In Table 2 we give a list of some reducible representation 
of  $SU(3)\times SU(2)\times U(1)$ that are anomaly free.
We observe that the unifying group $SU(5)$ predicts that matter fermions 
correspond to the choice of the last row. On the contrary, other rows are 
acceptable anomaly free representations that do not immediately lead to 
any unified group. 
\begin{table}
\begin{equation*}
\begin{tabular}{|l|l|l|l|}
\hline
observable & experimental value & SM\ prediction & pull \\
\hline 
$M_{Z}$ & $91.1876\pm 0.0021$ & $91.1874\pm 0.0021$ & 0.1 \\ 
$\Gamma _{Z}$ & $2.4952\pm 0.0023$ & $2.4968\pm 0.0011$ & -0.7 \\ 
$\sigma _{\text{had}}^{0}\left[ \text{nb}\right] $ & $41.541\pm 0.037$ & $%
41.467\pm 0.009$ & 2.0 \\ 
$R_{e}$ & $20.804\pm 0.050$ & $20.756\pm 0.011$ & 1.0 \\ 
$R_{\mu }$ & $20.785\pm 0.033$ & $20.756\pm 0.011$ & 0.9 \\ 
$R_{\tau }$ & $20.764\pm 0.045$ & $20.801\pm 0.011$ & -0.8 \\ 
$R_{b}$ & $0.21629\pm 0.00066$ & $0.21578\pm 0.00010$ & 0.8 \\ 
$R_{c}$ & $0.1721\pm 0.0030$ & $0.17230\pm 0.00004$ & -0.1 \\ 
$A_{FB}^{e}$ & $0.0145\pm 0.0025$ & $0.01622\pm 0.00025$ & -0.7 \\ 
$A_{FB}^{\mu }$ & $0.0169\pm 0.0013$ &  & 0.5 \\ 
$A_{FB}^{\tau }$ & $0.0188\pm 0.0017$ &  & 1.5 \\ 
$A_{FB}^{b}$ & $0.0992\pm 0.0016$ & $0.1031\pm 0.0008$ & -2.4 \\ 
$A_{FB}^{c}$ & 0.0707$\pm 0.0035$ & 0.0737$\pm 0.0006$ & -0.8 \\ 
$A_{FB}^{s}$ & 0.0976$\pm 0.0114$ & 0.1032$\pm 0.0008$ & -0.5 \\ 
$\bar{s}_{l}^{2}$ & 0.2324$\pm 0.0012$ & 0.23152$\pm 0.00014$ & 0.7 \\ 
& 0.2328$\pm 0.0050$ &  & -1.5 \\ 
$A_{e}$ & 0.15138$\pm 0.00216$ & 0.1471$\pm 0.0011$ & 2.0 \\ 
& 0.1544$\pm 0.0060$ &  & 1.2 \\ 
& 0.1498$\pm 0.0049$ &  & 0.6 \\ 
$A_{\mu }$ & 0.142$\pm 0.015$ &  & -0.3 \\ 
$A_{\tau }$ & 0.136$\pm 0.015$ &  & -0.7 \\ 
& $0.1439\pm 0.0043$ &  & -0.7 \\ 
$A_{b}$ & $0.923\pm 0.020$ & $0.9347\pm 0.0001$ & -0.6 \\ 
$A_{c}$ & $0.670\pm 0.027$ & $0.6678\pm 0.0005$ & 0.1 \\ 
$A_{s}$ & $0.895\pm 0.091$ & $0.9356\pm 0.0001$ & -0.4 \\ 
$M_{W}$ &  &  & \\
\hline
\end{tabular} 
\end{equation*}
\caption{The electroweak data and the Standard Model fit \cite{Erler:2006vt}.}
\end{table}
Without the assumption of a unified theory that includes 
a flavor symmetry, it remains a mystery why nature has chosen three times 
the last row  (Table 2) for the three families \cite{Georgi:1979md}. 
Also the Higgs hypercharge, that is a completely  arbitrary choice without
 unification hypotheses, find an obvious explanation within $SU(5)$.
Among all 
possible groups of unification SU(5), SO(10) and $E_6$ are the most
 favored 
candidates. These are the arguments in favor of unification, but we have not  
yet understood why the proton lifetime is  huge, if compared with the muon 
 decay and the neutron decay lifetime. 
This is the hierarchy problem,
 {\it i.e.}
the need of an explanation for the gauge lepto-quark  boson masses and the
 weak 
boson masses.
The effective potential responsible for the symmetry breaking pattern
 $SU(5)\rightarrow SU(3)\times
SU(2)\times U(1)\rightarrow SU(3)_{\text{col}}\times U(1)_{\text{em}}$ is
 written
\begin{equation}
V=-\mu ^{2}\,H^{2}-m^{2}\Sigma ^{2}+\lambda _{1}\,\ H^{4}+\lambda _{2}\,\
H\,\ \Sigma ^{2}\,H+\lambda _{3}\,\Sigma ^{4}    \label{eqpot}
\end{equation}
where $\Sigma $ and $H$ are respectively  the  {\bf 24} and the  {\bf 5} of
SU(5).
We have to choose the arbitrary parameters  $\mu ,m,\lambda_{2}$, with an
extreme fine tuning if we want the hierarchy 
$\tau_{\mathrm{prot}}\gg \tau_{\mu} $ between the proton and the muon
 lifetime.
We will see how it is possible to modify the standard theory in order to 
obtain a simple explanation of the hierarchy problem.
\begin{table}
\begin{equation*}
\begin{tabular}{|c|}
\hline
 (3,3)(-1)+(\={3},2)(4)+(\={3},1)(-5) \\  
\hline
(1,1)(-5/6)+(1,1)(-5/6)+(1,1)(-1/6)+(1,1)(1/3)+(1,1)(1/2)+(1,1)(1) \\ 
\hline
 (\={3},2)(4)+(3,\={2})(-4)+(1,2)(1)+(1,%
\={2})(-1) \ \ vectorlike \\ 
\hline
(1,\={2})(-1/2)+(%
\={3},1)(1/3)+(1,1)(1)+(\={3},1)(-2/3)+(3,2)(1/6)$\subset 10+\bar{5}$ \\%
\hline
\end{tabular}
\end{equation*}
\caption{Representations of the Standard Model gauge group 
SU(3)$\times$SU(2)$\times$U(1). The last row corresponds to the 10+$\bar{5}$, 
the minimal and anomaly free chiral representation of SU(5).}
\end{table}
\subsection{Non linear extension of quantum field theory}
The free classical hamiltonian of a scalar real field is written
\begin{equation}
\mathcal{H}=\int d^{3}\,x\,\,\pi ^{2}(x)+\phi (-\nabla ^{2}+m^{2})\phi (x).
 \label{cla}
\end{equation}
We have to replace the functions $%
\pi (x)$ and $\phi (x)$, defined in the three-dimensional space $x$,
 with two operators  $\hat{\pi}(x)$ and  $\hat{\phi}(x)$ that satisfy the 
commutation rules 
\begin{equation}
\left[ \hat{\pi}(x),\hat{\phi}(y)\right] =\,i\,\ \delta ^{3}(x-y). 
  \label{eqalg}
\end{equation}
This quantizes the  hamiltonian above (\ref{cla}).
We can also give a representation of the algebra (\ref{eqalg}) of the 
operators  $\hat{\pi}(x)$ and $\hat{\phi}(x)$ in the space of functionals
 $ S[\phi ]$, replacing  $\hat{\pi}(x)$ and  $\hat{\phi}(x)$ with 
\begin{equation}
\begin{tabular}{ll}
$\hat{\phi}(x)|S>$ & $\rightarrow \,\phi (x)\,S[\phi ]$ \\ 
$\hat{\pi}(x)|S>$ & $\rightarrow \,i\ \frac{\delta }{\delta \phi (x)}%
\,S[\phi ]$%
\end{tabular}
\end{equation}
It is easy to verify that they satisfy the algebra (\ref{eqalg})
\begin{equation}
\left[ i\,\frac{\delta }{\delta \phi (x)},\phi (y)\right] =\,i\,\ \delta
^{3}(x-y).
\end{equation}
In the  Schroedinger picture, the physical states of quantized field are 
described by the functional  $S[\phi ,t]$, whose time dependence $t$, is given
 by  the  Schroedinger equation  
\begin{equation}
i\,\frac{\partial }{\partial t}\,S[\phi ,t]=\int d^{3}x\,\left( -\ \frac{%
\delta ^{2}}{\delta \,\phi ^{2}(x)}\,-\phi (x)\nabla ^{2}\phi
(x)+m^{2}\,\phi ^{2}(x)\,\right) S[\phi ,t]   \label{eqscroe}
\end{equation}
The equation  (\ref{eqscroe}) represent the quantized theory of a free scalar
field. The mass $m$ is a fundamental and arbitrary constant. 
In the case of a free particle,  $m$ coincides with the 
 physical measured mass,
but in the general  case of an interacting field it does not coincide 
with the physical mass, because it also depends on the radiative corrections
 due   to the presence of interactions, and on 
any  possible vacuum expectation 
value  of other  scalar fields. 
For example in (\ref{eqpot}) the value of $\mu$ necessary to get a very light 
higgs at the weak scale, is around $10^{16}$ GeV, {\it i.e.} the order of 
magnitude of the  vev of $\Sigma$ . The  fine tuning is needed to achieve 
a cancellation between several contributions. In other words this  
correspond to a very precise choice for $\mu$, very close to the arrow   
depicted in Fig.1. 
Since $\mu$ is a free parameter, the choice of $\mu$ very close to the arrow 
(Fig.1) is accidental and would give not natural predictions.
Now we will see how this odd fine tuning can be explained in a non linear 
extension of the equation  (\ref{eqscroe}).
Let us assume that we add a non linear term that modifies  eq. (\ref{eqscroe})
as follows 
\begin{equation}
\begin{tabular}{rl}
$i\,\frac{\partial }{\partial t}\,S[\phi ,t]=$ & $ \hat{H}
 S[\phi ,t]+\int d^{3}x\,\ J(x,t)\,\ \phi ^{2}(x)\,\ S[\phi ,t]$
 \\ 
&\\
$J(x,t)=$ & $\int D\phi \,\ S^{\dagger }[\phi ,t]\,\phi (x)^{2}\,S[\phi ,t]$%
\end{tabular}
  \label{nonline}
\end{equation}
When the non linear term   $J(x)$ is very small and negligible the equation 
(\ref{nonline}) reduces to a linear equation and it describes an ordinary 
quantum field theory. But in certain physical situations $J(x)$ could be not 
negligible\footnote{When the physical state $S[\phi ,t]$ is a system that 
contains one (or more)scalar  particles $\phi$, then  $J$ is proportional to 
the wave function squared of this particle.}. 
Let us consider the case when  $J(x)$ is small but not negligible, and we can 
solve the equation  (\ref{nonline}) in perturbation theory.
The simplest non trivial case is when $J(x,t)$ is a constant and does not 
depend on space and time. This happens when the functional  $S$ corresponds
to physical systems where the field $\phi$ has constant and non zero vev.
For any fixed value of $J$ eq.(\ref{nonline}) is linear, and we know that 
such a linear equation admits a stationary solution $S[\phi,t]$ when  
the expectation value  of $\phi$ minimizes the effective potential
 (with $J$ fixed).
  $S[\phi,t]$ is the wave functional of the 
state with minimal energy.
The vev of $\phi$ depends on the arbitrary choice of $J$, but also $J$ 
(in the non linear case)  is a function of the vev $\phi$. 
Thus both the vev $\phi$ 
and the constant $J$ are two variables determined by two equations
 (\ref{nonline}).
The non linear term in (\ref{nonline}) can be replaced by 
 any generic dependence 
on the vev $\phi$, in fact   the  second eq.(\ref{nonline}) 
is an arbitrary physical choice. An illustrative  choice  like
 \begin{equation}
\mu^2 (\phi )=\mu^2+J=- M^2_{\text{unif}} \log(\phi /M_{\text{unif}})  
 \label{dipe}
\end{equation}
could even 
explain the hierarchy problem.
In fact in the linear theory the vacuum expectation value is a function 
of the arbitrary constant $\mu$ (see Fig.1), but in the non  linear theory 
$\mu$ is not arbitrary and 
depends on $\phi$ (see eq.(\ref{nonline}) and eq. (\ref{dipe})).
The special dependence (\ref{dipe}) explains why the intersection of both 
curves\footnote{The first curve is the dependence of the vev from $\mu$ as 
from the minimization of (\ref{eqpot}); the second curve comes from the 
dependence
 of $\mu$ (or equivalently $J$) from the vev.}
  (Fig.1) happens when the vev $\phi$ is very small {\it i.e.}
 close to the arrow.
This explains the hierarchy problem. 
\begin{figure}
\epsfig{figure=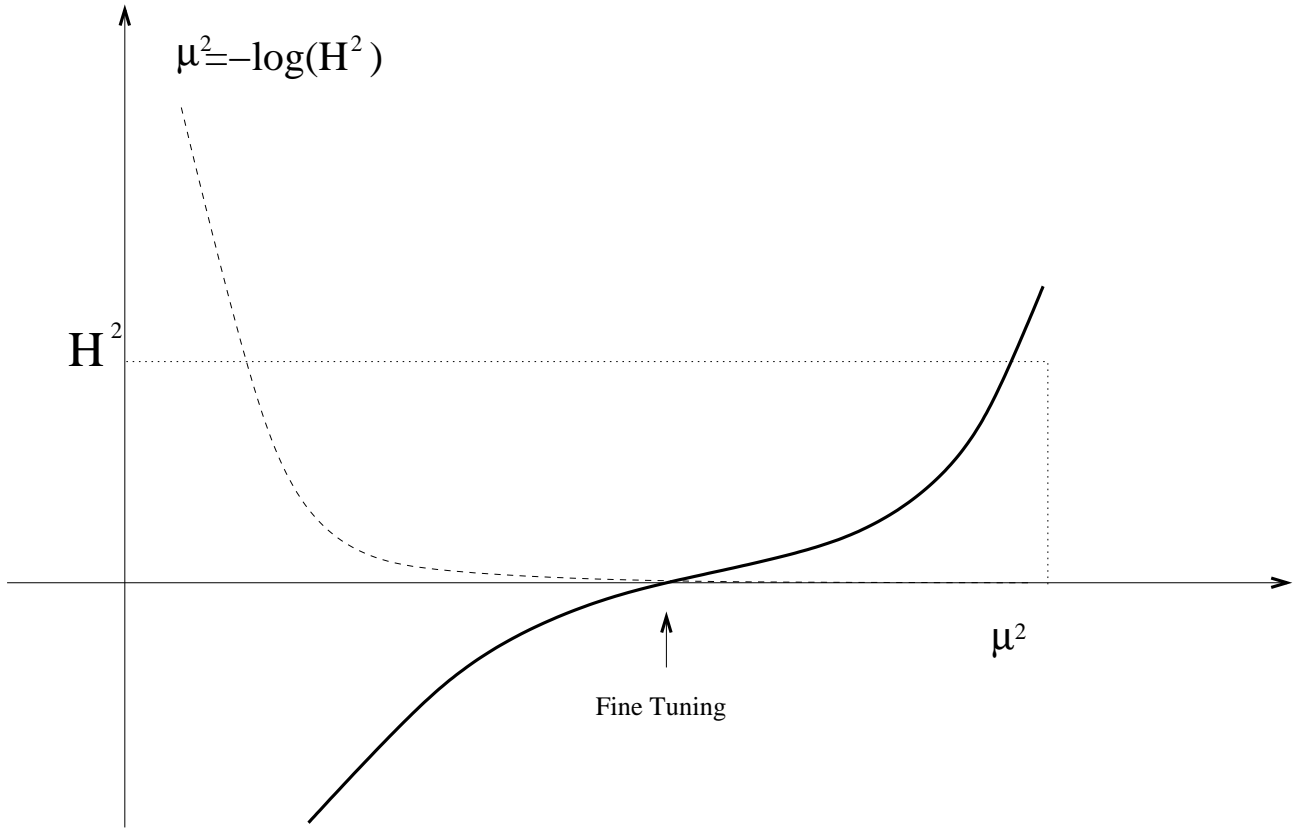,height=11cm}
\caption{ The Higgs doublet vev $H^2=<\phi^2>$ 
as a function of the bare mass $\mu^2$  (solid curve). The dashed curve comes 
from the non linear term and gives  
the bare mass   $\mu^2$ as a function of the vev $H$ (see eq.(\ref{dipe})
 in the text).}
\end{figure}

However a non linear extension of the  Schroedinger functional equation 
shows the lack of a theory of measurement.
If a state  $S,$  evolves 
from being the superimposition of several eigenstates toward 
a single eigenstate of an observable, because of a measurement,
 then this time evolution  also affects  (\ref{nonline}) 
and the probability distribution of the final states is automatically modified.
In other words the time evolution deduced from equation (\ref{nonline})
can be considered valid until when no measurement is performed\footnote{Note 
that even the definition of  measurement in quantum mechanics is rather 
ambiguous. 
And this put an ambiguity on the extent of validity of eq.(\ref{nonline}).}. 

There is another extension of the field theory that does not violate the
 quantum  mechanical principle of linear superimposition in the evolution 
of physical states and that could explain in a similar way the hierarchy 
problem. But before introducing this new theory we deepen briefly the safety 
of a collider like the LHC in the context of a non linear extension.  

It is not hard to realize that if we abandon the request of linearity in 
eq. (\ref{eqscroe}), various possible extensions are possible, each one 
with a phenomenology and with physical consequences
 that are completely  unexpected.
Even if an exhaustive discussion of all possible cases is very difficult or 
even impossible, we briefly draw our attention to few cases that probably 
deserve more attention.  
Firstly, let us note that the limit   
 \cite{Sher:1993mf,Isidori:2007vm} 
on the higgs mass due to the requirement of stability of 
the vacuum cannot be directly applied in a non  linear extension of the
 standard  theory. Let us now see some potential risks: the creation 
of a new exotic particle $\phi$ at the collider LHC locally 
changes the value of  $J$ (\ref{nonline}), that is  in the region  
occupied by the wave packet of this scalar particle. 
This could modify the fundamental bare constants of the linearized theory. 
It would also modify the physical masses and the couplings of the standard 
particles: for example the photon could become massive, and all
 electromagnetic interactions would be turned off in a region 
of finite volume\footnote{The theory of quantum mechanics does not put 
any bound on   the size of a wave packet.}. 

Another risk could come from the fact that the non linear theory  
violates the crossing symmetry and thus 
 the production of very light particles 
with strong interaction with matter is not incompatible with the observation 
of previous accelerators. We remind also that non linear interactions
with  the simultaneous presence of  significant amount of dark matter 
in the   solar systems adds other dangerous scenarios.

\subsection{Third Quantization}

A similar but  alternative explanation of the hierarchy problem is obtained 
embedding  second quantization into third quantization 
\cite{Caravaglios:2002ws,Caravaglios:2002qm}.
The embedding of first quantization into second quantization 
proceeds as follows.
The  Schroedinger equation for one particle is written
\begin{equation}
i\,\frac{\partial }{\partial t}\psi (x,t)=H\,\ \psi (x,t)
\label{eqscroe2}
\end{equation}
and in fact the quantum state of a particle in the  Schroedinger picture
is a wave function  $\psi (x).$ The wave function is replaced by 
 an operator when we  go to second quantization (quantum field theory)
\begin{equation}
\psi (x)\Rightarrow \hat{\psi}(x)
\end{equation}
and we set the following anticommutation rules
\begin{equation}
\left\{ \hat{\psi}(x),\hat{\psi}^{\dagger }(y)\right\} =\delta ^{3}(x-y).
\end{equation}
The quantum field theory analogue of eq.(\ref{eqscroe2}) is eq.(\ref{eqscroe}).
This equation   (\ref{eqscroe}) tells us that the quantum state of the
 universe  is described 
by a functional  $S[\phi ,t]$ where the variable  $t$ denotes the 
time evolution of the physical state. If we repeat the same steps as for  
going 
from first quantization to second quantization, and we want to embed 
second quantization into third quantization, then the functional  $S[\phi ]$
becomes an operator
\begin{equation}
S[\phi ]\Rightarrow \hat{S}[\phi ]
\end{equation}
that satisfies the anticommutation rules 
\begin{equation}
\left\{ \hat{S}[\phi ],\hat{S}^{\dagger }[\phi ^{\prime }]\right\} =\delta
(\phi -\phi ^{\prime }).  \label{com}
\end{equation}
As an illustrative example,  the simplest hamiltonian can be written
  
\begin{equation}
\mathcal{H}=\int D\phi \,\ d^{3}x\,\ \hat{S}^{\dagger }[\phi ]\left( -\ 
\frac{\delta ^{2}}{\delta \,\phi ^{2}(x)}\,-\phi (x)\nabla ^{2}\phi
(x)+m^{2}\,\phi ^{2}(x)\,\right) \hat{S}[\phi ].    \label{terza}
\end{equation}
The vacuum state  $|0>$ satisfies the condition  
\begin{equation}
\mathcal{H}|0>=0  
\end{equation}
and represents a state without  fields and without space, while the state 
\begin{equation}
|F>=\int D\phi \,\ F[\phi ]\,\ \ \hat{S}^{\dagger }[\phi ]\,|0>
\end{equation}
with 
\begin{equation}
F[\phi ]=\,\ \exp (-\frac{1}{2}\int d^{3}x\,\ \phi (x)\sqrt{-\nabla
^{2}+m^{2}}\phi (x))  \label{minimo}
\end{equation}
represents the state of a universe with only one scalar field $\phi $
and with minimal energy. 
It is not difficult to verify the the functional   (\ref{minimo}) minimizes 
the energy $E$ among all possible states   $|F>$ 
\begin{equation}
E=<F|\,\mathcal{H\,}|F>.
\end{equation}
Let us see why such a theory can explain the hierarchy problem. 

We can add to the hamiltonian  (\ref{terza}) new composite 
operators that contain 
a larger number of creation/annihilation  $\hat{S},\hat{S}^{\dagger }$
 operators. We add to the hamiltonian  $\mathcal{H}$ the following interaction 
\begin{equation}
\mathcal{H_{\mathrm{int}}}=\int D\phi ~d^3 x \sum_{i=1}^n 
 a_n \hat{S}^{\dagger }[\phi_1 ]\cdots 
\hat{S}^{\dagger }[\phi_n ] \phi_1^2(x)\cdots  \phi_n^2(x)
\hat{S}[\phi_1 ]\cdots \hat{S}[\phi_n ].
\end{equation}
We introduce the function $G(J)$ defined by the sequence of $a_n$ as follows 
\begin{equation}
G(J)=\sum_{n=1}^\infty a_n J^n
\end{equation}
We have a considerable freedom in the  $G(J)$, and almost any choice of 
$G(J)$  corresponds to  a physically 
acceptable\footnote{Unfortunately we have not yet 
(in third quantization) a highly constraining theoretical principle 
 such  ``renormalizability'', that  applies only in second quantization.
 Thus we have a lot of freedom in this embedding and in the choice of 
 $\mathcal{H}_{\mathrm{int}}$.}   $\mathcal{H}_{\mathrm{int}}$.
 In those cases where one can apply the mean field approximation, the vacuum 
does not satisfy the trivial relation 
\begin{equation}
S[\phi ]\,\ |0>=0.
\end{equation}
On the contrary, the action of several annihilation operators   $S$ is 
the following
\begin{equation}
\hat{S}[\phi _{1}]\,\hat{S}[\phi _{2}]\ \cdot \cdot \cdot \hat{S}[\phi
_{n}]|0>\simeq F[\phi _{1}]F[\phi _{2}]\cdot \cdot \cdot F[\phi _{n}]|x>
\end{equation}
where $F$ is a functional that must be determined by the minimization 
of $E$ 
\begin{equation}
E=<0|\,\mathcal{H\,}|0>
\end{equation}
that leads us  to the equation
\begin{equation}
\left( -\ \frac{\delta ^{2}}{\delta \,\phi ^{2}(x)}\,-\phi (x)\nabla
^{2}\phi (x)+(m^{2}\,+G(J))\;\phi ^{2}(x)\,+\gamma \,\phi ^{4}\right) F[\phi
]=\lambda \,\ F[\phi ].    \label{fond}
\end{equation}
where $J$ is given by 
\begin{equation}
J=\int D\phi \,\ F^{\dagger }[\phi ]\,\ \phi ^{2}(x)\,\ F[\phi ]. 
\label{vev}
\end{equation}
The equation  (\ref{fond}) is not linear in $F$ but it can be solved 
as follows. Firstly, let us neglect eq.(\ref{vev}), and let us  
assume that $J$ is an arbitrary constant (an external source) 
that does not depend on $F$.
With this assumption, the equation  (\ref{fond})  is much more simple, since
 it is linear and we know how to solve it, by means of ordinary quantum field 
theory methods. In fact the eq.(\ref{fond}) is the same equation that we solve 
to find out the state with minimal energy (the vacuum) in quantum field theory,
we have to calculate and 
 minimize the effective potential where $G(J)$ appears as an external source: 
it corrects the bare mass with the replacement
\begin{equation} 
m^2\rightarrow  m^2+G(J)\equiv \mu^2.  \label{jdip}
\end{equation}
The field  $\phi $ takes a vev if $\mu^2 \equiv  m^2+G(J)$ is negative; the vev will 
be a function of $J$, through the dependence (\ref{jdip}). But also $J$ is a 
function   of the vev as predicted by the original exact equation (\ref{vev}).
We have two variables and two equations: both the vev $\phi$ and $J$ are 
determined. This clearly appears in Fig. 1, where the solid curve gives
the dependence of the vev on the $\mu^2$, as predicted by the minimization of
 the full effective potential ({\it i.e.} including all radiative 
corrections). The dashed  curve gives the dependence of $\mu^2$ on $J$, where we 
have assumed a logarithmic function for $G(J)$. In this case the intersection 
of the two curves occurs very close to the arrow: it is not a fine tuned
and arbitrary choice, the hierarchy  is enforced by the 
 logarithmic function $G(J)$.     

This theory of third quantization has another interesting 
 direct prediction, concerning the flavor problem: it provides us with an 
explanation for the existence of fermion families.
We have already mentioned that the existence of three fermion families 
with quantum numbers given by the last row in Table 1, hints a group 
of unification beyond the Standard Model. However the grand unified theory
does not  tell us why there are three identical families.
In the past several unifying group have been studied, in the attempt to 
understand the three families. No convincing and significant result has been 
found. In third quantization our universe (made of three identical fermion 
 families) is obtained applying three consecutive times
 the creation operator $S^\dagger[\psi]$ on the vacuum state
\begin{equation}
\int D \psi~ F[\psi_1,\psi_2,\psi_3] \,
S^\dagger[\psi_1]\,S^\dagger[\psi_2]\,S^\dagger[\psi_3]\,|0>.
\end{equation}  
The functional $F$ identifies the physical quantum state of our universe, and 
the three functions  $\psi_i$ represent the fermionic fields of the three 
families.
In the general case the functional operators  $S^\dagger$ create new families, 
and we can call them family creation operators.
The anticommutation rules (\ref{com}) tell us that the functional $F$ is 
antisymmetric when we exchange the fields  $\psi_i$, not only at $t=0$, but 
for the full time evolution: the hamiltonian of second quantization that 
describes  the time evolution of $F$ must be symmetric under permutations
of the fermions  $\psi_i$. 

We have obtained 
 a simple explanation of the family problem and a clear prediction 
on the flavor symmetry group.
Namely the flavor symmetry is the permutation group $S_n$ where $n$ is the 
number of families. We still have to understand if the functional $S$ only 
depends on the fermion field  $\psi$ or it is preferable to add 
the dependence on  the gauge boson $A^\mu$ too: in the last case the operator  
 $S^\dagger[\psi,A^\mu]$ creates universe containing $n$ families, with the 
following   gauge group and flavor symmetry 
\cite{Caravaglios:2006aq,Babu:2007mb}
\begin{equation}
G^n\, > \hskip-6pt \lhd \,  S_n    \label{gru}
\end{equation}
where $G$ is a unified gauge group  and the permutations  $S_n$ act both 
on the fermionic families  and the gauge bosons families, exchanging the $n$
factors in the group   $G^n$. It remains to understand which gauge  group 
$G$ to choose. SU(5) is a possible group \cite{Haba:2005ds} but it is a 
symmetry that 
does not automatically contain 
righthanded neutrinos ({\it i.e.}  gauge bosons ignore the 
righthanded  neutrinos): we have not explored this possibility.
SO(10) is the most appealing  candidate\cite{Babu:2007mb} ,
 because it contains the righthanded neutrino in the {\bf 16}.
In the simplest $SO(10)$ model where the higgs doublet is in the {\bf 10},
we  have yukawa unification between the dirac neutrino masses and the up quark 
masses. This must be discarded. There are interesting exceptions 
 to this unification if we put 
the Higgs into larger irreducible representations but  this 
study is left for another work.

We have decided to focus on the gauge group $E_6$.
Differently from SO(10), whose {\bf 16} contains only one Standard Model 
singlet, the   {\bf 27} of $E_6$ contains two singlets of the Standard Model
  ($SU(3)\times SU(2)\times U(1) $). The lefthanded neutrino of the standard 
model can exchange  a yukawa  interaction with both singlets. 
While for  the first singlet, precisely as in SO(10), this interaction 
coincides  with the  yukawa interaction in the up quark sector, the 
 coupling between the second singlet and the lefthanded neutrino does 
not unify with other yukawa  fermion couplings.
Namely, the scalar representation  {\bf 351$^\prime$} of $E_6$
contains various  $SU(2)$ doublets with different quantum numbers, and a  
particular one   gives a yukawa interaction for neutrinos only
\begin{equation}
\lambda\, 27\, 27\, 351^\prime=\lambda\, \nu_L \nu_R \, H   
\label{yuk}
\end{equation}
while all other fermions contained in the {\bf 27} have a combination of 
quantum numbers such that any yukawa coupling with the Higgs doublet in 
(\ref{yuk}) is forbidden.
The interaction  (\ref{yuk}) allows us to understand why the neutrino 
Dirac mass does not unify  with up quark mass.
After having chosen the group  $G=E_6$ we must fix the number $n$ in
  (\ref{gru}).
The simplest and more obvious choice is $n=3$, but this choice does not 
help us in understanding why the two almost degenerate states 
(the first two columns in (\ref{pmns})) in neutrino
 oscillations are the  $S_3$ singlet and the component of the $S_3$ doublet 
that is even under the exchange of the two heaviest families 
(the $S_2$ symmetry). In other words the mass hierarchy between the even   
states  and the odd state  under  $S_2$ 
suggests a $S_2$  symmetry 
and not $S_3$; but we need  $n\ge 3$ in (\ref{gru}) if we want (at least)
three families. In fact even in those cases  $n>3$,
some pattern of symmetry breaking of the group  (\ref{gru}) lead us 
to the Standard Model with three families of fermionic matter.
It is just in these cases that we also find an explanation for neutrino
 masses and mixing as observed in neutrino oscillations. 
For clarity, we study the case $n=4$, because the generalization to the case 
with arbitrary  $n>3$ is trivial. Our aim is to explain how to attain in 
neutrino oscillations the  mixing angle matrix \cite{Fogli:2005gs}
\begin{equation}
\begin{pmatrix}
\begin{tabular}{ccc}
$\frac{-2}{\sqrt{6}} $ & $\frac{1}{\sqrt{3}} $ & 0 \\
$\frac{1}{\sqrt{6}} $ & $\frac{1}{\sqrt{3}} $ &$\frac{-1}{\sqrt{2}} $  \\
$\frac{1}{\sqrt{6}} $ & $\frac{1}{\sqrt{3}} $ & $\frac{1}{\sqrt{2}} $ \\
\end{tabular}
\end{pmatrix}  \label{pmns}
\end{equation}
with  $\Delta m^2_{\mathrm{atm}}\gg\Delta m^2_{\mathrm{sol}}$.
We have three distinct possibilities: the neutrino  mass matrix is diagonal 
and the oscillations are due to an off-diagonal  charged lepton mass matrix.
The second case is when the   charged lepton mass matrix is diagonal.
The last possibility is when both matrices are not diagonal.

We will assume that the lepton charged matrix is diagonal, thus the columns of 
the matrix ({\ref{pmns}) coincide with the three mass eigenstates of neutrinos
in the flavour basis. They are also eigenstates of the symmetry $S_2$ 
that exchanges the last two rows in the  ({\ref{pmns}).
The second column is a singlet of the $S_3$ symmetry that permutes the rows.

In the following model we will try to explain the matrix (\ref{pmns}), and why 
 $|\Delta m^2_{\mathrm{atm}}|\gg|\Delta m^2_{\mathrm{sol}}|$, but we 
will ignore  the sign of  $ \Delta m^2_{\mathrm{sol}}$, because it requires a 
more detailed study.
The  $S_n$ symmetry  ($n\ge 3$) can hardly explain the pattern 
  $m_3^2\gg m_1^2 = m_2^2$, but it can more easily explain why  
\begin{equation}
m_3^2\gg m_1^2\gg m_2^2.   \label{gera}
\end{equation}
In fact, the seesaw mechanism   changes the (\ref{gera}) into 
 $m^R_{\mathrm{sing}}\gg m^R_{\mathrm{doub}}$:
 the righthanded $S_3$ singlet becomes 
the heaviest state. So the  $S_3$ symmetric righthanded neutrino matrix must 
be of the form 
\begin{equation}
M_\nu^R\simeq\begin{pmatrix}
\begin{tabular}{ccc}
$m_d$&$ m $&$ m$\\
$m$& $m_d$ & $m$\\
$m$&$ m$ & $m_d$\\
\end{tabular}
\end{pmatrix}.  \label{sing}
\end{equation}
with 
\begin{equation}
m_d=m. \label{id}
\end{equation}
The matrix (\ref{sing}) descends
 from the $S_3$ symmetry, while 
eq. (\ref{id}) does not.  The reason why the $S_3$ doublet is much more  light
 is obscure.

If we add a fourth family, we can write the following antisymmetric matrix 
\begin{equation}
M=\begin{pmatrix}
\begin{tabular}{cccc}
0&0&0&-1\\
0&0&0&-1\\
0&0&0&-1\\
1&1&1&0 \\
\end{tabular}
\end{pmatrix}  \label{anti}
\end{equation}
that has the following properties: it is $S_3$ symmetric, {\it i.e.}
 it is invariant 
under the exchange of the first three families. It couples only with 
$S_3$ singlets, the only states acquiring a non zero mass. 
The doublet of  $S_3$ is given by the two massless states 
$(-2/\sqrt{6},1/\sqrt{6},1/\sqrt{6},0)$ and 
 $(0,-1/\sqrt{2},1/\sqrt{2},0)$.

 The matrix (\ref{anti}) is the only  4$\times$4  matrix that is 
simultaneously $S_3$ symmetric and antisymmetric under transposition.
Instead of majorana masses, 
we are forced to choose a dirac mass term
\begin{equation}
 M_{ij}\, \nu^i_R\, X^j_R   \label{dir}
\end{equation} 
with two distinct weyl spinors $\nu_R$ and $X_R$, 
otherwise the   (\ref{dir}) would be identically zero, since $M_{ij}=-M_{ji}$.
The  {\bf 27} of $E_6$ contains two different weyl spinors, that 
we can call   $\nu_R$ and $X_R$;  thus  (\ref{anti}) and  (\ref{dir}) 
 are  compatible 
with the choice of the group  $E_6^4\, > \hskip-6pt \lhd\, S_4 $.

We complete this discussion, suggesting  how to break the  group  $S_3$
into $S_2$. We add a  scalar field  $\phi^i$, with the family index 
$i=1,4$. Only the first component of this field takes a vev  $\phi^1=v$.
The state $(-2/\sqrt{6},1/\sqrt{6},1/\sqrt{6},0)$ takes a mass, while the 
orthogonal state  $(0,-1/\sqrt{2},1/\sqrt{2},0)$ remains as the lightest 
righthanded neutrino. The seesaw mechanism through the diagonal yukawa 
interaction (\ref{yuk}) will make the $S_2$ odd state
 (last column of (\ref{pmns})) the heaviest neutrino. A more detailed 
discussion of this model can be found in \cite{Caravaglios:2006aq}.


\begin{thebibliography}{99}

\bibitem{Chanowitz:2008ix}
  M.~S.~Chanowitz,
  arXiv:0806.0890.

\bibitem{Sher:1993mf}
  M.~Sher,
  Phys.\ Lett.\  B {\bf 317} (1993) 159
  ,Addendum-ibid.\  B {\bf 331} (1994) 448

\bibitem{Isidori:2007vm}
  G.~Isidori, V.~S.~Rychkov, A.~Strumia and N.~Tetradis,
  Phys.\ Rev.\  D {\bf 77} (2008) 025034



\bibitem{Georgi:1979md}
  H.~Georgi,
  Nucl.\ Phys.\  B {\bf 156} (1979) 126.


\bibitem{Erler:2006vt}
  J.~Erler,
  AIP Conf.\ Proc.\  {\bf 857} (2006) 85


\bibitem{Caravaglios:2002ws}
  F.~Caravaglios,
  arXiv:hep-ph/0211183.

\bibitem{Caravaglios:2002qm}
  F.~Caravaglios,
  arXiv:hep-ph/0211129.



\bibitem{Caravaglios:2006aq}
  F.~Caravaglios and S.~Morisi,
  Int.\ J.\ Mod.\ Phys.\  A {\bf 22} (2007) 2469


\bibitem{Babu:2007mb}
  K.~S.~Babu, S.~M.~Barr and I.~Gogoladze,
  Phys.\ Lett.\  B {\bf 661} (2008) 124


\bibitem{Haba:2005ds}
  N.~Haba and K.~Yoshioka,
  Nucl.\ Phys.\  B {\bf 739} (2006) 254


\bibitem{Fogli:2005gs}
  G.~L.~Fogli, E.~Lisi, A.~Marrone, A.~Palazzo and A.~M.~Rotunno,
   arXiv:hep-ph/0506307.

\end{thebibliography}
\end{document}